# *Excited state characterization of carbonyl containing carotenoids: a comparison between single and multireference descriptions*

Riccardo Spezia[a*], Stefan Knecht[b*] and Benedetta Mennucci[c*]

[a]LAMBE, Université d'Evry Val d'Essonne, CEA, CNRS, Université Paris Saclay, F-91025 Evry (France).
[b]Laboratorium für Physikalische Chemie, ETH Zürich, Vladimir-Prelog-Weg 2, 8093 Zürich, Switzerland
[c]Dipartimento di Chimica e Chimica Industriale, Università di Pisa, via G. Moruzzi 13, 56124 Pisa, Italy
[†]

**Abstract**

Carotenoids can play multiple roles in biological photoreceptors thanks to their rich photophysics. In the present work, we have investigated six of the most common carbonyl containing carotenoids: Echinenone, Canthaxanthin, Astaxanthin, Fucoxanthin, Capsanthin and Capsorubin. Their excitation properties are investigated by means of a hybrid density functional theory (DFT) and multireference configuration interaction (MRCI) approach to elucidate the role of the carbonyl group: the bright transition is of $\pi\pi^*$ character, as expected, but the presence of a C=O moiety reduces the energy of $n\pi^*$ transitions which may become closer to the $\pi\pi^*$ transition, in particular as the conjugation chain decreases. This can be related to the presence of a low-lying charge transfer state typical of short carbonyl-containing carotenoids. The DFT/MRCI results are finally used to benchmark single-reference time-dependent DFT-based methods: among the investigated functionals, the meta-GGA (and in particular M11L and MN12L) functionals show to perform the best for all six investigated systems.

# 1. Introduction

Carotenoids are among the most important natural pigments due the multiple roles that they play in biological systems. For example, in light harvesting pigment-protein complexes they can act both as efficient antenna in the mid-visible range of the solar spectrum or as photoprotection agents.[1,2] Such a multiplicity of roles is made possible by the richness and flexibility of their electronic states. Due to their conjugated structure, in fact, carotenoids present different ππ* states whose nature and energy ordering is largely dependent on the conjugation length. For a qualitative description of these states, it is common to adopt an analysis based on a $C_{2h}$ symmetric picture. According to this picture, the lowest singlet excited state presents the same ($A_g$) symmetry of the ground state and, as a result of the selection rules, the corresponding transition is forbidden. On the contrary, the transition to the second singlet excited state is allowed, as its symmetry is now $B_u$.

This symmetry-based analysis has been widely applied in the interpretation of the photophysical features of many carotenoids even when they have a much lower symmetry due to additional groups capping the conjugated chain or placed within the chain. In many observations, however, this picture has shown some limits and the presence of additional intermediates states have been suggested to achieve a more complete explanation of spectroscopic evidences.[1,3,4] In particular, in the case of carbonyl-containing carotenoids the study of solvation effect on their photophysics has raised the hypothesis that a low-lying internal charge transfer (ICT) state can be populated.[5,6,7,8,9,10] Some authors have also suggested a connection between ICT and nπ* states.[11,12,13]

Unfortunately, an exhaustive theoretical description of the electronic nature of carotenoids is not straightforward due to many different reasons. First of all, the large dimensions of these systems have prevented (and still do) the application of very accurate quantum-mechanical (QM) methods. Secondly, the wide range of possible systems, both in terms of different conjugation length, and different atomic composition, has made a systematic study almost impossible. Moreover, the high geometrical flexibility of the conjugated chain makes a theoretical analysis even less straightforward as a single calculation on the minimum geometry could not be representative of what is observed for the system in solution or within a biomacromolecular matrix. All these aspects are finally combined with the further difficulty that the ππ* states of carotenoids can exhibit a double excitation character which changes with the composition of the system and/or the presence of side groups. A proper QM description can thus not be solely based on a single reference method if an unambiguous picture of the relative position and the nature of the various ππ* states is sought for. Among suitable QM methods, the one that has shown the best performances is the combination of a Density Functional Theory (DFT) description with a multi-reference configuration interaction (MRCI) ansatz.[14,15,16] This method has been applied to study excitations of carotenoids by several groups.[17,18,19,20,21] More recently some of the authors have exploited the DFT/MRCI approach to investigate the ππ* excitation of different carotenoids present in the light-harvesting pigment-protein complexes of plants and algae.[22,23] These studies clearly showed that, in all considered cases, contributions from multiple excitations are crucial for the description of the first ($S_1$) (and third, $S_3$) excited states, whereas the second ($S_2$) state is dominated by single excitations. In particular, for the longer carotenoids, a considerable mixing of the $S_2$ and $S_3$ states was found with a corresponding decrease of the excitation energy and a parallel increase of the transition dipole of the $S_3$ state.

Here, the same approach is employed to achieve a better understanding of the unique (and often still puzzling) photophysics of a specific class of carotenoids, namely those containing carbonyl groups. In particular, we have studied

Echinenone, Canthaxanthin, Astaxanthin (in both keto and enol forms), Fucoxanthin, Capsanthin and Capsorubin. The chosen pigments have different functions and are found in different biological systems. In particular, Echinenone is the pigment involved in the photoactivity of the Orange Carotenoid Protein (OCP) which is responsible of photoprotection of cyanobacteria.[24] Canthaxanthin has a structure very similar to Echinenone (see **Figure 1**) and is a red-orange pigment found in different organisms (such as crustaceans, fishes and algae). Astaxanthin is responsible for the coloration of lobsters and other blue-black crustaceans.[25] Fucoxanthin is the pigment present in light-harvesting complexes involved in efficient marine photosynthesis.[26] Finally, Capsanthin and Capsorubin are responsible for the color of red pepper and paprika.

The detailed photophysics and photochemistry of each system is beyond the scope of the present paper; each one should constitute *per se* a specific study. Here, instead, we focused on Franck-Condon transitions of the relevant conformers for each carotenoid with the goal of understanding the real nature of the involved electronic states. To achieve such a goal, we made use of the multireference DFT/MRCI approach. Moreover, the DFT/MRCI results were employed as a benchmark for the single-reference Time-Dependent (TD) DFT formulation. Using TD-DFT is particularly intriguing for such systems because of its computational efficiency. Previous analyses have shown that the Tamm–Dancoff approximation (TDA) of TD-DFT in combination with meta-GGA functionals give a generally good description of the transition properties of the three lowest states.[27] This behavior can be traced back to error cancelations originating from the differential effects that a neglect of the multideterminant character has on the ground and excited states:[27] the importance of double excitations grows faster for the ground state than for the excited states, and the result is that the errors caused by the neglect of double excitations in the ground and excited states will match at some point. Following along this line in the present study, we draw a comparison between DFT/MRCI and TDA-DFT with the aim of identifying the class of functionals which provides best correct transition properties and state orderings.

By analyzing the character of the transitions, we elucidate how the conjugation length of the carotenoid influences the relative position of the (dark) $n\pi^*$ transitions arising from the presence of the C=O group with respect to the (bright) $\pi\pi^*$ ones: increasing the conjugation length the separation between the two states increases. The excited states ordering (and their transition energies), characterized with DFT/MRCI is best reproduced by the meta-GGA M11L and MN12L functions.

The paper is organized as follows. In Section 2 we introduce the theoretical methods employed in this work, followed by results on transition energies and excited states ordering as obtained by both DFT/MRCI and TDA-DFT (providing the necessary data to pinpoint the best functionals in connection with the TDA-DFT approach). In Section 4, we analyze the nature of the main characteristic excited states and in Section 5 we summarize and give an outlook to future work.

## 2. Methods

### 2.1 MRCI calculations

Single-point multireference DFT/MRCI calculations[14,15] at the respective DFT-optimized structures (*see below*) have been performed employing the recently re-parametrized DFT/MRCI Hamiltonian.[16] Following our established computational recipe of previous works[22,28], we employ a split valence basis set (def2-SV(P))[29] for all atom types. Optimized Kohn-Sham (KS) orbitals using the BH-LYP functional[48,30] as implemented in the Turbomole 6.3 program package[31] constitute the basis for the construction of configuration state functions in the subsequent MRCI expansion. With the 1s electrons of the heavy atoms kept frozen in the post-SCF steps, the initial MRCI reference space was spanned by all single and double excitations from the six highest occupied MOs to the five lowest unoccupied MOs of the ground state KS determinant. In this first step we calculate the twelve lowest eigenvectors with an orbital selection threshold set to 0.6 $E_h$. The new reference space for the second DFT/MRCI step then comprised all configurations contributing to one of the twelve lowest-lying eigenvectors of the initial DFT/MRCI calculation with a squared coefficient of 0.003 and larger. We then optimized the ground- and lowest seven electronically excited states of the target carotenoid where the orbital selection threshold was increased to at least 1.0 $E_h$ (the actual threshold is determined dynamically and may be up to 1.2 $E_h$).

**2.2 DFT calculations**

All ground-state carotenoid structures were optimized with DFT/B3LYP and a 6-311G(d,p) one-particle basis set. The starting structures for Echinenone and Canthaxanthin were taken from a recent work [32] while for Astaxanthin we extracted the structure from the X-ray data of β-Crustacyanin where this pigment is present (PDB code 1GKA) and added the missing H atoms prior to the optimization. For the other carotenoids we built the corresponding starting structures based on what is reported in previous calculations on their structure[33,34]. Following the structural optimization we performed a vibrational analysis to verify that the obtained stationary points correspond to minimum energy structures on the potential energy surface.

As previously reported for echinenone and canthaxanthin,[32] DFT/B3LYP geometries are able to correctly reproduce available vibrational spectroscopic information, in particular $v_1$ and $v_2$ resonance Raman peaks, as shown in **Table S1** of the ESI. The resulting good agreement with experimental vibrational data gives us confidence on the correctness of the structures optimized with DFT/B3LYP. All xyz coordinates of the newly optimized structures are collected in the ESI. With the optimized ground-state structures at hand, we then performed TDA-DFT calculation [35] to characterize the first four vertical exited states. Calculations on Echinenone and Canthaxanthin served as our initial benchmark of DFT functionals with respect to predicting correctly the ordering of excited states, focusing foremost on the oscillator strength associated with each electronic transition. To this end we compiled a set of different functionals following the main classes as follows:

(i) GGA: BLYP[36,37,38] and PBE[39];
(ii) meta-GGA: TPSS,[40] RevTPSS,[41,42] VSXC,[43] HCTH,[44] M06L,[45] M11L,[46] MN12L[47];
(iii) hybrid-GGA: B3LYP,[38,48] PW91,[49] BHandH PBE0[50];
(iv) hybrid-meta-GGA: TPSSh[51] and M06[52,53];
(v) range separated hybrid-GGA: HSEH1PBE,[54,55] CAM-B3LYP[56] and LC-ωPBE.[57,58]

In all cases a 6-311+G(d,p) basis set was used.
Based on the performance of the above functionals with respect to the excitation energies and properties of Echinenone and Canthaxanthin (see **Tables S2-S6** of the ESI) we chose a subset of functionals, namely TPSS, M06L, M11L and MN12L, and in addition the popular B3LYP functional, to further characterize the excited states of the other carbonyl-containing carotenoids (Astaxanthin, Fucoxanthin, Capsanthin and Capsorubin).
All TDA-DFT calculations were performed with the Gaussian 09 software.[59]

**3. Comparison between DFT/MRCI and TDA-DFT**

We first investigated the lowest four excited states of Echinenone and Canthaxanthin by comparing DFT/MRCI with TDA-DFT with the five selected functionals. Although the full s-cis structures (CC) are the most stable ones (see details in Ref. 32), we considered in addition the structures resulting from an s-cis/trans isomerization (TC, CT and TT) (see **Figure 1**).
The calculated excitation energies for the four lowest transitions are shown in **Table 1** whereas in **Figures 2 and 3** we report the corresponding oscillator strengths.
The DFT/MRCI data confirm what is already known for other carotenoids, and polyenes in general, concerning the lowest-lying excited states: the $S_2$ state, which is of typical ππ* character, is the first bright state, while $S_1$, $S_3$ and $S_4$ are (in general) dark states. We postpone here a more detailed discussion of the character of the different electronically excited states to Section 4.

|   | n | n_r | n_o | n_tot |
|---|---|-----|-----|-------|
| a) | 9 | 1$^T$+1$^C$ | 1$^T$ | 12 |
| b) | 9 | 2$^C$ | 1$^C$ | 12 |
| c) | 9 | 2$^T$ | 1$^T$ | 12 |
| d) | 9 | 1$^C$+1$^T$ | 1$^C$ | 12 |
| e) | 9 | 2$^T$ | 2$^T$ | 13 |
| f) | 9 | 1$^C$+1$^T$ | 1$^C$+1$^T$ | 13 |
| g) | 9 | 2$^C$ | 2$^C$ | 13 |

**Figure 1.** Echinenone and Canthaxanthin isomers studied: (a) T(rans)C(is), (b) CC, (c) TT, (d) CT Echinenone, and (e) TT, (f) CT, and (g) CC Canthaxanthin. The conjugation lengths are also reported. n: C=C bond in polyene chain, $n_r$: C=C bond linked to polyene chain in the rings; $n_o$: C=O bond connected to the polyene chain and/or to the $n_r$. $n_{tot} = n + n_r + n_o$. C and T superscripts denote the cis and trans connection, respectively.

|  | Echinenone | | | | Cantaxanthin | | | |
|---|---|---|---|---|---|---|---|---|
|  | $S_1$ | $S_2$ | $S_3$ | $S_4$ | $S_1$ | $S_2$ | $S_3$ | $S_4$ |
| CC | | | | | | | | |
| B3LYP | 2.27 | 2.64 | 3.06 | 3.18 | 2.27 | 2.53 | 3.10 | 3.19 |
| TPSS | 1.89 | 2.15 | 2.26 | 2.46 | 1.95 | 2.04 | 2.09 | 2.10 |
| M06L | 1.96 | 2.29 | 2.39 | 2.56 | 2.04 | 2.11 | 2.29 | 2.30 |
| M11L | 1.92 | 2.21 | 2.49 | 2.52 | 1.99 | 2.05 | 2.43 | 2.43 |
| MN12L | 2.06 | 2.36 | 2.69 | 2.80 | 2.15 | 2.16 | 2.71 | 2.72 |
| **MRCI** | **2.12** | **2.40** | **2.54** | **2.71** | **2.01** | **2.34** | **2.60** | **2.64** |
| TC | | | | | | | | |
| B3LYP | 2.21 | 2.56 | 3.01 | 3.09 | 2.22 | 2.45 | 3.04 | 3.05 |
| TPSS | 1.83 | 2.13 | 2.19 | 2.44 | 1.88 | 2.01 | 2.03 | 2.04 |
| M06L | 1.90 | 2.26 | 2.33 | 2.53 | 1.96 | 2.08 | 2.23 | 2.24 |
| M11L | 1.86 | 2.17 | 2.46 | 2.47 | 1.92 | 2.02 | 2.37 | 2.38 |
| MN12L | 2.00 | 2.31 | 2.66 | 2.75 | 2.06 | 2.13 | 2.65 | 2.66 |
| **MRCI** | **1.96** | **2.32** | **2.50** | **2.55** | **1.92** | **2.29** | **2.42** | **2.49** |
| CT | | | | | | | | |
| B3LYP | 2.21 | 2.56 | 3.01 | 2.24 | | | | |
| TPSS | 1.84 | 2.17 | 2.22 | 2.44 | | | | |
| M06L | 1.91 | 2.25 | 2.41 | 2.54 | | | | |
| M11L | 1.87 | 2.16 | 2.48 | 2.55 | | | | |
| MN12L | 2.00 | 2.30 | 2.67 | 2.84 | | | | |
| **MRCI** | **1.99** | **2.25** | **2.46** | **2.97** | | | | |
| TT | | | | | | | | |
| B3LYP | 2.13 | 2.50 | 2.96 | 3.09 | 2.18 | 2.38 | 2.98 | 3.04 |
| TPSS | 1.76 | 2.13 | 2.15 | 2.41 | 1.82 | 1.98 | 1.98 | 1.98 |
| M06L | 1.83 | 2.21 | 2.33 | 2.50 | 1.90 | 2.03 | 2.18 | 2.18 |
| M11L | 1.80 | 2.12 | 2.44 | 2.47 | 1.86 | 1.97 | 2.33 | 2.33 |
| MN12L | 1.92 | 2.26 | 2.63 | 2.74 | 2.00 | 2.08 | 2.60 | 2.61 |
| **MRCI** | **1.86** | **2.17** | **2.31** | **2.81** | **1.86** | **2.15** | **2.25** | **2.76** |
| EXP | 2.56-2.44 | | | | 2.50-2.31 | | | |

**Table 1.** DFT/MRCI and TDA-DFT transition energies (in eV) for the first four excited states of Echinenone (CC, TC, CT and TT) and Canthaxanthin (CC, TC and TT). Experimental data are taken from Ref. 32: for Echinenone we report the values corresponding to OCP-o and OCP-r, and the range obtained for different isomers, while for Canthaxanthin we report the energy value range from experiments in different solvents.

Moving to TDA-DFT results, as expected, B3LYP can be used only to obtain information on the bright state while the energy ordering cannot be reproduced. On the contrary, meta-GGA functionals give a much better description.[22] This is indeed the case for Canthaxantin, for which all meta-GGA functionals report dark $S_1$, $S_3$ and $S_4$ states (oscillator strength zero or very small) and a bright $S_2$ state. For Echinenone, instead, the outcome is less straightforward to interpret, as the $S_1$ state oscillator strength is generally comparable to that of the $S_2$ state. Notably, the TPSS functional, that was previously reported to provide correct excited state ordering in other carotenoids,[22] inverts the $S_1$ and $S_2$ states (as well as $S_3$ and $S_4$). M06L, M11L and MN12L, on the other hand, reproduce better the oscillator strengths ordering, while still providing a too large oscillator strength for $S_1$.

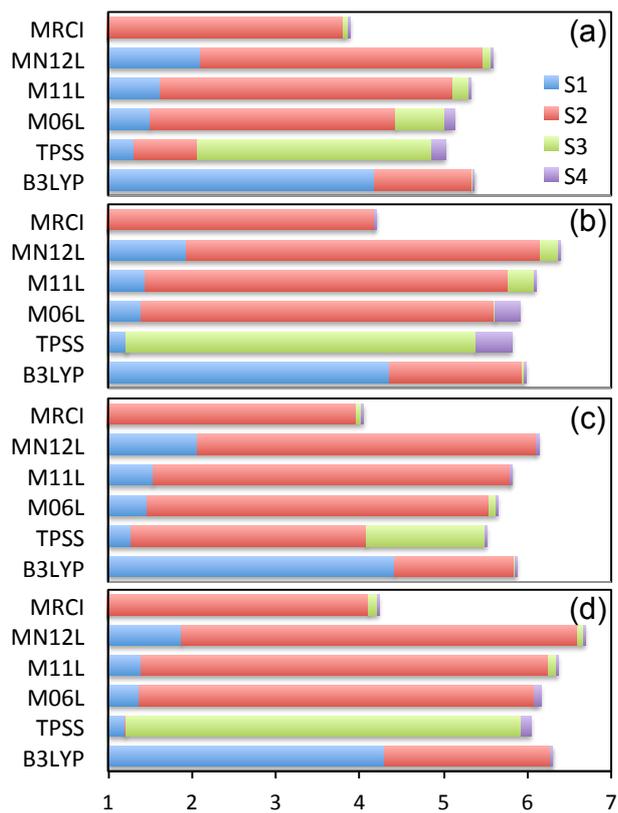

**Figure 2.** Comparison between DFT/MRCI and TDA-DFT oscillator strengths of the first four excited states (results in a.u.) : a) CC-Echinenone; b) TC-Echinenone; c) CT-Echinenone; d) TT-Echinenone

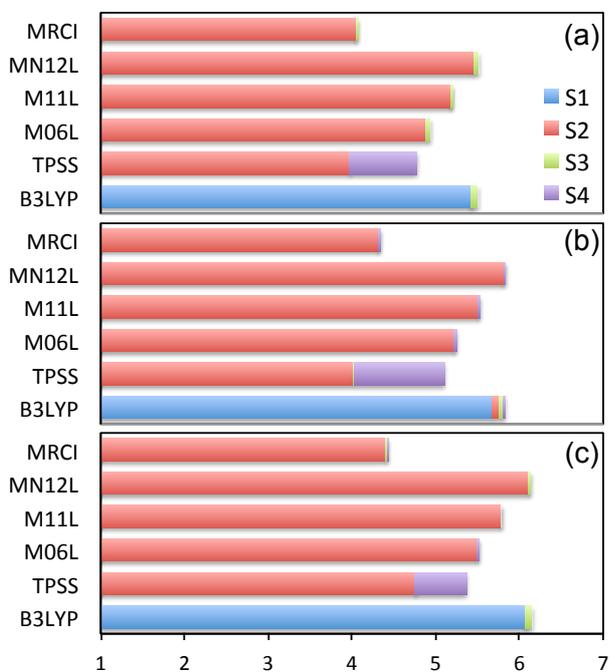

**Figure 3.** Comparison between DFT/MRCI and TDA-DFT oscillator strengths of the first four excited states (results in a.u.) : a) CC-Canthaxanthin; b) TC-Canthaxanthin; c) TT-Canthaxanthin

One important property of Echinenone (and presumably of other carotenoids) is the difference in light absorption energy due to s-cis/trans isomerization.[32, 60] We have thus compared the transition energy differences between bright states upon different isomerization pathways, as summarized in **Table 2**.

|          | Echinenone |         |        |
|----------|------------|---------|--------|
|          | CC-TC      | CC-CT   | CC-TT  |
| B3LYP    | 0.065      | 0.064   | 0.138  |
| TPSS     | 0.061      | 0.030   | 0.101  |
| M06L     | 0.028      | 0.038   | 0.072  |
| M11L     | 0.038      | 0.048   | 0.082  |
| MN12L    | 0.050      | 0.058   | 0.101  |
| DFT/MRCI | 0.081      | 0.074   | 0.147  |
| Experiments | 0.14 (2.48-2.34) | | |

**Table 2.** Energy differences (in eV) for isomerizations of transitions to bright states ($S_1$ for B3LYP, $S_3$ in Echinenone for TPSS and $S_2$ for M06, M11L, MN12L and DFT/MRCI). Experimental data are from Ref. 32 and they correspond to the difference between OCP-o and OCP-r (in parenthesis the values of electronic transition energies).

Experimentally, an estimation of 0.14 eV difference in light absorption is reported,[32] in agreement with CC-to-TT isomerization as obtained by DFT/MRCI and B3LYP (note that for B3LYP we consider the $S_0$-$S_1$ transition energies). Among all meta-GGA functionals comprised in our test set, MN12L reproduces these values best. We note in passing that Takano and co-workers reported in a recent work [61] excited state data for 3'-hydroxyechinenone (with the H-atom at the 3' position of TC-Echinenone substituted by a hydroxyl group) calculated with TD-DFT/CAM-B3LYP as well as RASSCF/RASPT2 and a 6-31G(d) basis set. They calculated excited state energy difference upon isomerization for the $S_2$ state are within the range of 0.02-0.03 eV for the s-trans to s-cis isomerization that is to be compared with the CC-TC Echinenone isomerization data summarized in **Table 2**. Considering the good agreement of our DFT/MRCI data with the available experimental reference data for the unsubstituted echinenone isomers, the excited state energy differences due to isomerization obtained by Takano and co-workers for 3'-hydroxyechinenone are probably too low.

Taking into account all the properties investigated so far (excited state ordering, transition dipoles and energy differences upon different s-cis/trans isomerizations) we conclude that MN12L is the best meta-GGA functional among those considered here and that other Minnesota meta-GGA functionals work generally well. In contrast, the other meta-GGAs may lead to erroneous conclusions, as shown by the incorrect prediction of oscillator strengths of echinenone associated with the respective electronic transitions.

The performances of the same meta-functionals have been further investigated on the excited states of other carbonyl containing carotenoids: Astaxanthin (in both keto and enol forms), Fucoxanthin, Capsanthin and Capsorubin (see **Figure 4**). As before, we have also considered the B3LYP functional since, albeit providing the wrong state ordering, it should yield transition energies in good agreement with experiments for the bright state.

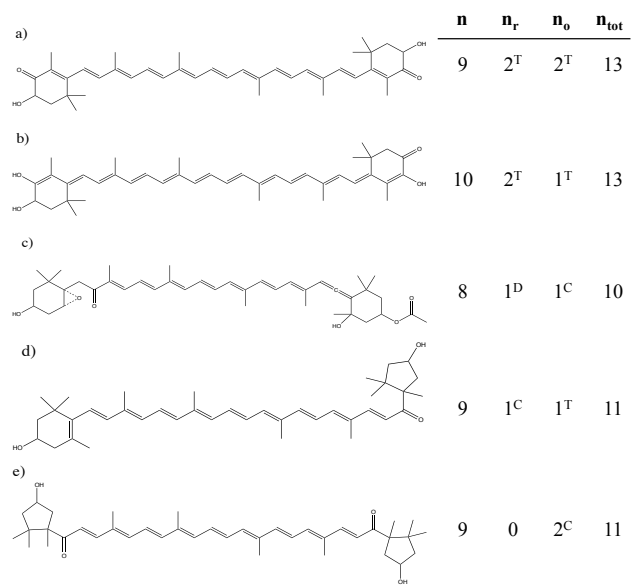

**Figure 4.** Other carbonyl containing carotenoids: (a) Astaxanthin (keto), (b) Astaxanthin (enol), (c) Fucoxanthin, (d) Capsanthin and (e) Capsorubin. The conjugation lengths are also reported. n: C=C bond in polyene chain; $n_r$: C=C bond linked to polyene chain in the rings; $n_o$ : C=O bond connected to the polyene chain and/or to the $n_r$. $n_{tot}$= n + $n_r$ + $n_o$. C and T superscripts denote the cis and trans connection, respectively. For Fucoxanthin D superscript stays for the diene moiety linked to the polyene chain.

All results are summarized in **Table 2** and **Figure 5** where we compare transition energies and oscillator strengths obtained with either approach.

| | $S_1$ | $S_2$ | $S_3$ | $S_4$ |
|---|---|---|---|---|
| | *Astaxanthin-keto* | | | |
| B3LYP | 2.14 | 2.33 | 2.94 | 3.14 |
| TPSS | 1.78 | 1.93 | 2.23 | 2.24 |
| M06L | 1.85 | 1.98 | 2.40 | 2.42 |
| M11L | 1.82 | 1.94 | 2.39 | 2.52 |
| MN12L | 1.96 | 2.04 | 2.57 | 2.71 |
| **DFT/MRCI** | **1.80** | **2.10** | **2.20** | **2.71** |
| | *Astaxanthin-enol* | | | |
| B3LYP | 1.97 | 2.32 | 2.74 | 2.88 |
| TPSS | 1.51 | 1.95 | 2.15 | 2.28 |
| M06L | 1.59 | 2.01 | 2.25 | 2.36 |
| M11L | 1.57 | 1.95 | 2.22 | 2.32 |
| MN12L | 1.68 | 2.08 | 2.38 | 2.48 |
| **DFT/MRCI** | **1.64** | **2.03** | **2.11** | **2.48** |
| EXP | 2.33[a] | | | |
| | *Fucoxanthin* | | | |
| B3LYP | 2.56 | 2.97 | 3.19 | 3.66 |
| TPSS | 2.25 | 2.31 | 2.60 | 2.86 |
| M06L | 2.33 | 2.52 | 2.66 | 3.06 |
| M11L | 2.29 | 2.56 | 2.66 | 3.01 |
| MN12L | 2.43 | 2.73 | 2.94 | 3.16 |
| **DFT/MRCI** | **2.38** | **2.56** | **2.90** | **3.01** |
| EXP | 2.40[b]; 2.74[b] | | | |
| | *Capsanthin* | | | |
| B3LYP | 2.31 | 2.72 | 3.05 | 3.18 |
| TPSS | 1.95 | 2.10 | 2.37 | 2.59 |
| M06L | 2.02 | 2.28 | 2.43 | 2.68 |
| M11L | 1.94 | 2.33 | 2.43 | 2.62 |
| MN12L | 2.13 | 2.48 | 2.71 | 2.82 |
| **DFT/MRCI** | **2.09** | **2.34** | **2.63** | **2.76** |
| EXP | 2.64[c] | | | |
| | *Capsorubin* | | | |
| B3LYP | 2.43 | 2.64 | 2.99 | 2.99 |
| TPSS | 1.95 | 1.95 | 2.10 | 2.27 |
| M06L | 2.13 | 2.14 | 2.19 | 2.32 |
| M11L | 2.13 | 2.25 | 2.30 | 2.30 |
| MN12L | 2.30 | 2.37 | 2.59 | 2.59 |
| **DFT/MRCI** | **2.11** | **2.36** | **2.62** | **2.71** |
| EXP | 2.64[c] | | | |

**Table 2.** TDA-DFT and DFT/MRCI transition energies (in eV) for the first four excited states of Astaxanthin, Fucoxanthin, Capsanthin and Capsorubin. [a]Experiments in CHCl$_3$ from Ref. 62; [b]Experiments in THF from Ref. 33: a dark and a bright state were reported and assigned here to $S_1$ and $S_2$ based on our DFT/MRCI results. [c] Experiments in hexane from Ref. 63.

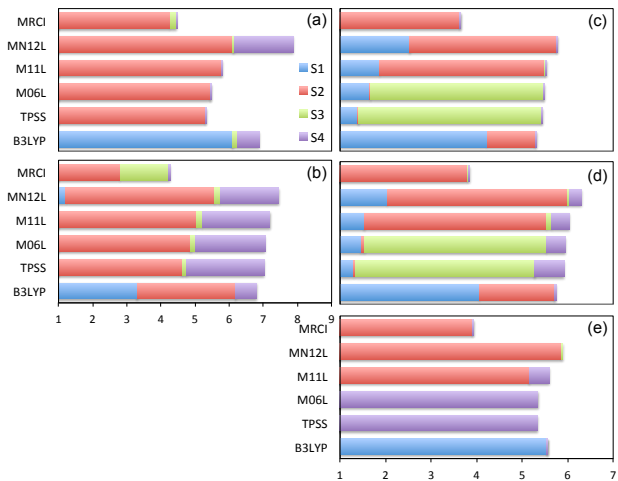
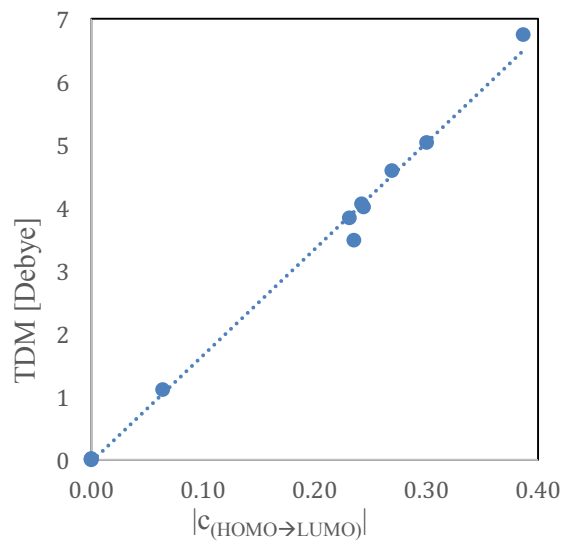

**Figure 5.** Comparison between DFT/MRCI and TDA-DFT oscillator strengths of the first four singlet excited states ($S_1$-$S_4$): (a) Astaxanthin/keto form, (b) Astaxanthin/enol form, (c) Fucoxanthin, (d) Capsanthin and (e) Capsorubin.

From the DFT/MRCI results we note that the $S_2$ state is the first bright state for all four carotenoids considered here. Interestingly, though, the $S_1$ ($S_3$) state in Fucoxanthin/Capsanthin and Astaxanthin, respectively, borrows some intensity (with oscillator strength in the range of 0.2-0.4 and up to 1.4 for the $S_3$ state of Astaxanthin in enol form). This intensity borrowing in the $S_1$ ($S_3$) state seems to be intimately connected with the amount of configuration mixing between doubly- and singly excited configurations, in particular the $\pi\pi^*$ (HOMO→LUMO) singly-excited configuration, that is the dominant configuration of the bright ($S_2$) state. To corroborate this hypothesis we show in **Figure 6** the transition dipole moment of the $S_0$→$S_1$ transition as a function of the absolute value of the CI coefficient $c_{(HOMO\rightarrow LUMO)}$ in the DFT/MRCI wave function expansion of the $S_1$ state.

**Figure 6.** Module of the transition dipole moment (TDM) of the $S_0$→$S_1$ electronic transition (in Debye) as a function of the absolute magnitude of the CI coefficient of the (HOMO→LUMO) single excitation as obtained from DFT/MRCI calculations for our set of carbonyl-containing carotenoids. $S_0$→$S1$ transitions with an absolute coefficient $c_{(HOMO\rightarrow LUMO)}$ less than 0.003 are not listed. The dotted line indicates a linear regression fit of the present data with an $R^2$ value of 0.995.

Clearly, a transition dipole moment for the $S_0$→$S_1$ transition that deviates considerably from 0 goes hand in hand with a $|c_{(HOMO\rightarrow LUMO)}|$ coefficient greater than 0.2. As a result, the largest coefficients (about 0.3-0.4) refer to the $S_1$ states of Fucoxanthin and Capsanthin, respectively. A similar analysis for the CI wave function composition of the $S_3$ state in Astaxanthin reveals $c_{(HOMO\rightarrow LUMO)}$ coefficients of 0.14 and 0.46 for the keto and enol form, respectively. Notably, in the enol form, the (HOMO→LUMO) singly-excited configuration is even the dominant configuration in the DFT/MRCI wave function expansion of the $S_3$ state.

Marian *et al.* have shown that for quinoid carotenoids,[64] additional dark states, predominantly with $n\pi^*$ character originating from excitations of the oxygen lone pair electrons, are energetically below the bright $\pi\pi^*$-like state. These predictions were based on DFT/MRCI calculations employing the original parameterization of the DFT/MRCI Hamiltonian.[14] Since these findings appear to be odd at first glance, in **Table S7** we compare excitation data obtained with the originally parameterized DFT/MRCI Hamiltonian[14] to those obtained with the very recently re-parameterized DFT/MRCI Hamiltonian.[16] In agreement with Lyskov *et al.*,[16] we observe that the original DFT/MRCI parameterization leads to a general underestimation of $n\pi^*$ excitation energies ultimately resulting in an upshift of the bright state to $S_3$ (Fucoxanthin and Capsanthin) and $S_4$ (Capsorubin) whereas the $n\pi^*$ states are predicted to be the $S_2$ state in Fucoxanthin (2.44 eV) and Capsanthin (2.30 eV), respectively, as well as to be a pair of degenerate $n\pi^*$ states ($S_2/S_3$, 2.35 eV) in Capsorubin. By contrast, with the new parametrization,[16] that we consider as our reference, we restore the "expected" excited-state picture with the $S_2$ state being always the first bright state among the four carotenoids as can be seen from **Table S7** and **Figure 5**, respectively. While the bright ($\pi\pi^*$) transition is consistently shifted towards lower energies by up to 0.11 eV (Astaxanthin[enol]), the $n\pi^*$ transitions are in turn shifted toward higher transition energies leading to the "expected" state ordering. For example, for Fucoxanthin the $n\pi^*$ transition becomes the third excited state ($S_3$, 2.90 eV) while in all remaining carotenoids the corresponding $n\pi^*$ transition(s) are located among the $S_4$-$S_6$ singlet-excited electronic states. In particular, we now find the degenerate pair of $n\pi^*$ transitions in Capsorubin as $S_4/S_5$ states with a transition energy of 2.71 eV and the $n\pi^*$ transition in Capsanthin at 2.76 eV ($S_4$).

Considering next the corresponding TDA-DFT data, in the case of Astaxanthin, all meta-functionals provide the expected excited-state ordering, whereas B3LYP predicts as bright state the first excited state $S_1$. Concerning the excitation energies, all TDA-DFT calculations yield transition energies similar to the DFT/MRCI reference data, with MN12L giving the best agreement, in particular for the bright state. Compared to experimental data, all calculations agree rather well with the reported 0-0 transition of 2.33 eV,[62] which was estimated from a broad absorption band. More recently, Begum *et al.*[65] reported a value of 2.61 eV obtained in $H_2O$/MeOH mixture in which there should be a keto-enol equilibrium.

In the case of Fucoxanthin and Capsanthin, the DFT/MRCI calculations yield the $S_2$ as the bright state at 2.56 eV and 2.34 eV, respectively. This particular feature is captured by the M11L and MN12L functionals. On the contrary, M06L and TPSS predict the $S_3$ as the bright state and hybrid-GGA B3LYP functional gives, consistently with the case of Astaxanthin, the $S_1$ state as the bright state. Compared to experiments[33,63] DFT/MRCI underestimates the excitation energy to the $S_2$ state in Fucoxanthin by about 0.2 eV while the excitation energy to the dark state agrees well with the estimation from van Grondelle and co-workers.[33]

Finally, for Capsorubin, TPSS and M06L wrongly locate the bright state as the fourth excited singlet state ($S_4$), while M11L and MN12L correctly locate it as $S_2$. Moreover, the corresponding TDA-DFT/B3LYP calculation predicts the $S_1$ state as bright state. Experimental data available[63] are about 0.15 eV higher than DFT/MRCI results.

Given the overall performances of the functionals in catching the correct bright state and the excitation energies as compared to the DFT/MRCI data, we conclude that M11L and MN12L are the most reliable functionals. They are able to correctly locate the bright state as $S_2$ in all carotenoids considered in this work, while other meta-functionals, like M06L and TPSS, tend to provide a wrong order in some of the present cases. Concerning excitation energies, of all DFT functionals comprised in our test set, M11L yields the best estimates of the transition energies in comparison with DFT/MRCI, although MN12L performs similarly well.

## 4. The nature of the excited states

In light of the results discussed above, we further analyzed the electronic character of the low-lying singlet excited states of the set of carbonyl-containing carotenoids. It is useful in this context to take into account not only the conjugation length of the polyene backbone but also the additional (conjugated) C=C and C=O double bonds that can be either in a cis or trans configuration with respect to the chain. To this end, we report in **Figures 1** and **4** the conjugation lengths of each carotenoid considered in this work as a function of the above parameters, namely (i) the length of the polyene conjugation (n), (ii) the number of C=C double bond(s) ($n_r$) connected to the polyene chain, (iii) the number of C=O double bond(s) ($n_o$) connected to the polyene chain and/or to the C=C double bond(s), as well as (iv) the total conjugation length $n_{tot} = n + n_r + n_o$.

Starting with an analysis of the DFT/MRCI data, we first note that the excitation energies of the first "dark" state ($S_1$) as well as of the "bright" state ($S_2$) are clearly correlated with the total conjugation length $n_{tot}$ of the respective carotenoid. The longer the total conjugation chain the lower the respective transition energies which could have been not only expected from a simple particle-in-a-box model but is also consistent with earlier findings for non-carbonyl containing carotenoids.[22] Accordingly, the $S_0 \rightarrow S_1$ and $S_0 \rightarrow S_2$ transition energies are largest for Fucoxanthin ($n_{tot}$=10) whereas they reach their minimum values in the enol form of Astaxanthin ($n_{tot}$=13). Although the keto form of Astaxanthin formally has the same total number of conjugated double bonds as the enol form, the presence of an additional trans-conjugated C=O double bond leads to an increase of the transition energies. Moreover, we can conclude that for the isomers of Echinenone (Canthaxanthin) the transition energies of the lowest-lying singlet states are predominantly a function of $n + n_r$ and to a lesser extent of $n_o$, as a closer look at their conjugation paths reveals. If, for a given isomer, the number $n_r$ of trans-connected C=C double bonds is larger than for another isomer of the same molecule, the transition energies will be lower. If, however, they are equal then we expect the excitation energies of $S_1$ and $S_2$ to be similar. For example, for Echinenone-TC and Echinenone-CT we find for the $S_1$ ($S_2$) states transition energies of 1.96 eV (2.24 eV) and 1.99 eV (2.25 eV), respectively, along with similar oscillator strengths for both transitions as shown in **Figures 2b** and **2c**. The latter is also reflected in the CI coefficients for the (HOMO→LUMO) single- and (HOMO→LUMO)$^2$ double-excitation configuration of the DFT/MRCI wave functions which for the $S_1$ ($S_2$) state is 0.23-0.24 (0.85) and 0.43 (0.11), respectively, in either case.

Turning next to the nπ* transitions, which are of single-excitation nature (CI coefficients corresponding to double excitation configurations are in all cases smaller than 0.2), their relative transition energy with respect to the valence ππ* transitions, in particular to the $S_1$ and $S_2$ states, correlates well with the conjugation pattern (see **Figure 7**).

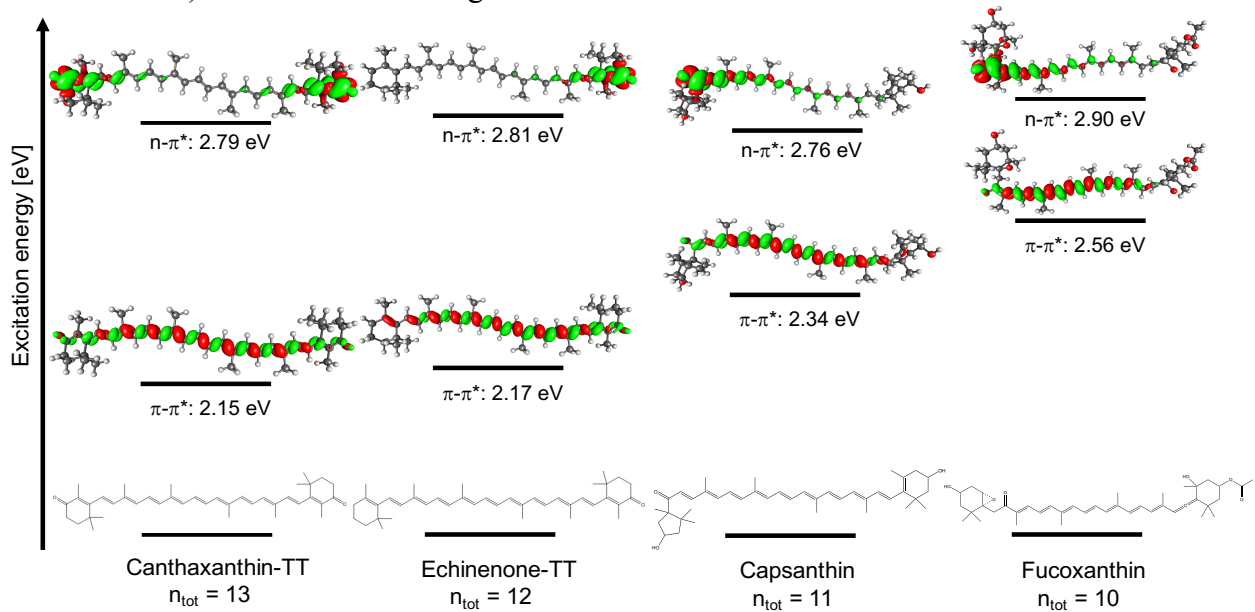

**Figure 7**. Energy gap between the (bright) ππ* and (dark) nπ* transitions as a function of conjugation length. All data obtained with DFT/MRCI. In the difference density plots (isosurface value of 0.001) green (red) indicates a negative (positive) density difference between the ground- and the respective electronically excited singlet state.

It is also interesting to note the dependence on the conformation. Considering for example the CC and TC isomers of Echinenone, we identify in all cases the $S_4$ state as the corresponding $n\pi^*$ transition but a significant lowering of the energy gap to the valence $\pi\pi^*$ transitions by about 0.2 eV is found when moving from the cis to the trans configuration for the CO bond. On the contrary, the transition energies of the valence $\pi\pi^*$ transitions remain comparable for both isomers. Similar considerations hold also for the isomers of Canthaxanthin where the energetic gap of the (degenerate) $n\pi^*$ transitions with respect to the $S_2$ state decreases from 0.7 eV in Canthaxanthin-CC ($\Delta(S_4/S_5 - S_2)$) to 0.6 eV in Canthaxanthin-TT ($\Delta(S_5/S_6 - S_2)$).

In order to better understand the TDA-DFT characterization of the excited states with respect to the DFT/MRCI analysis, in **Figure 8** we report the correlation between $|c_{(HOMO \rightarrow LUMO)}|$ and the oscillator strengths for M11L and M06L (for simplicity we report only these two functionals as representative of meta-GGA results).

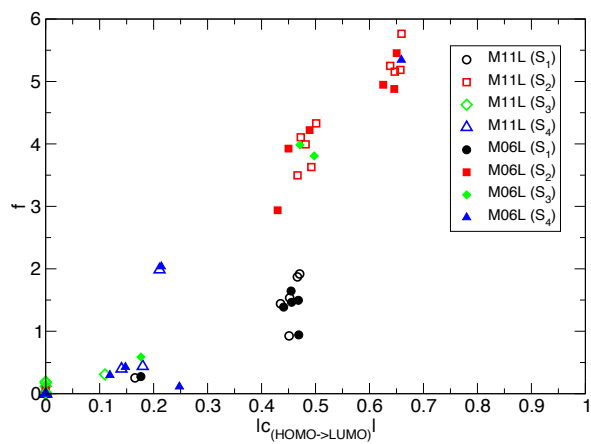

**Figure 8.** Oscillator strength (f) of the $S_0 \rightarrow S_n$ electronic transition as a function of the absolute magnitude of the CI coefficient of the (HOMO→LUMO) transition as obtained from TDA-M11L and TDA-M06L calculations.

In case of M11L, the transition to $S_2$ has always $|c_{(HOMO \rightarrow LUMO)}|$ larger than 0.45 and thus the transition dipole moment is the largest. On the other hand, for M06L in some cases transitions to $S_3$ and $S_4$ have a large $|c_{(HOMO \rightarrow LUMO)}|$ coefficient and they become the bright state.

We have to note that because of the single excitation character of the $n\pi^*$ transitions, they are well described by TDA-DFT. On the contrary, the double excitation character of the lowest $\pi\pi^*$ transition, necessarily entails inaccuracies in connection with the predictability of the TDA-DFT approach. This particularly holds when the HOMO-LUMO character of the transition (which in TDA-DFT inherently has only single excitation character) increases. As a result, by reducing the conjugation length, the transition energies for the $\pi\pi^*$ states not only go up but they are also overestimated which leads, for some functionals, to state crossings with $n\pi^*$ transitions. This is the reason why, for example, M06L in Fucoxanthin (which has the lowest number of conjugated bonds $n+n_r$) predicts the $S_3$ as the bright state (see **Figure 5c**).

As a final analysis, we compare ground-to-excited state density plot difference for the four lowest transitions obtained from DFT/MRCI calculations with natural transition orbitals (NTO)[66] calculated with TDA-DFT. We have chosen Fucoxantin and Capsantin as prototypical examples and reported results obtained for M06L (in which there are inversions between electronic transitions) and M11L (which provides a correct state ordering). From a comparison of the corresponding plots of **Figures 9** and **10**, it is evident that M11L qualitatively reproduces the DFT/MRCI description for the three lowest transitions of both systems: the two lowest transitions are both $\pi\pi^*$ transitions while the third is a clear $n\pi^*$ transition. We only note that for $S_1$ a much larger involvement of the 6-ring is found with respect to the DFT/MRCI description where it is instead localized in the central chain.

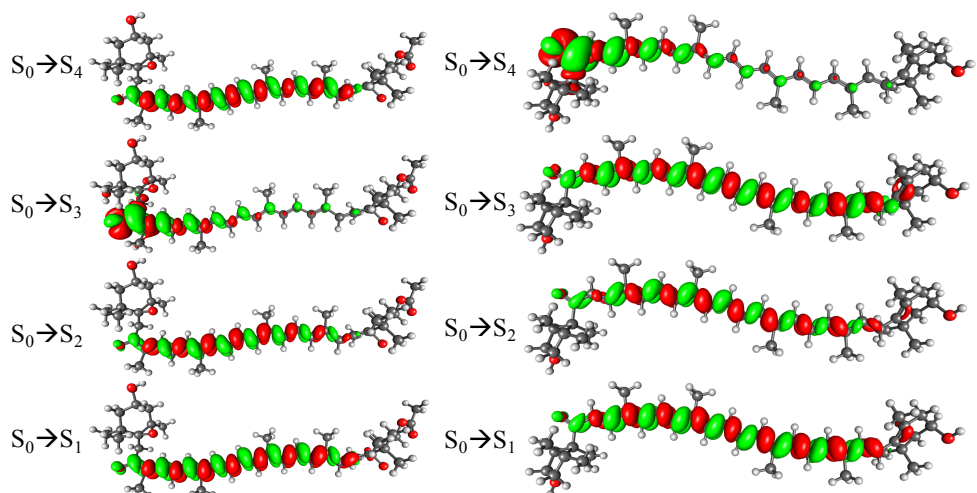

**Figure 9**. Comparison of difference density plots (isosurface value of 0.001) of the four lowest transitions in Fucoxanthin (left) and Capsanthin (right) obtained from DFT/MRCI calculations. Green (red) indicates a negative (positive) density difference between the ground- and the respective electronically excited singlet state.

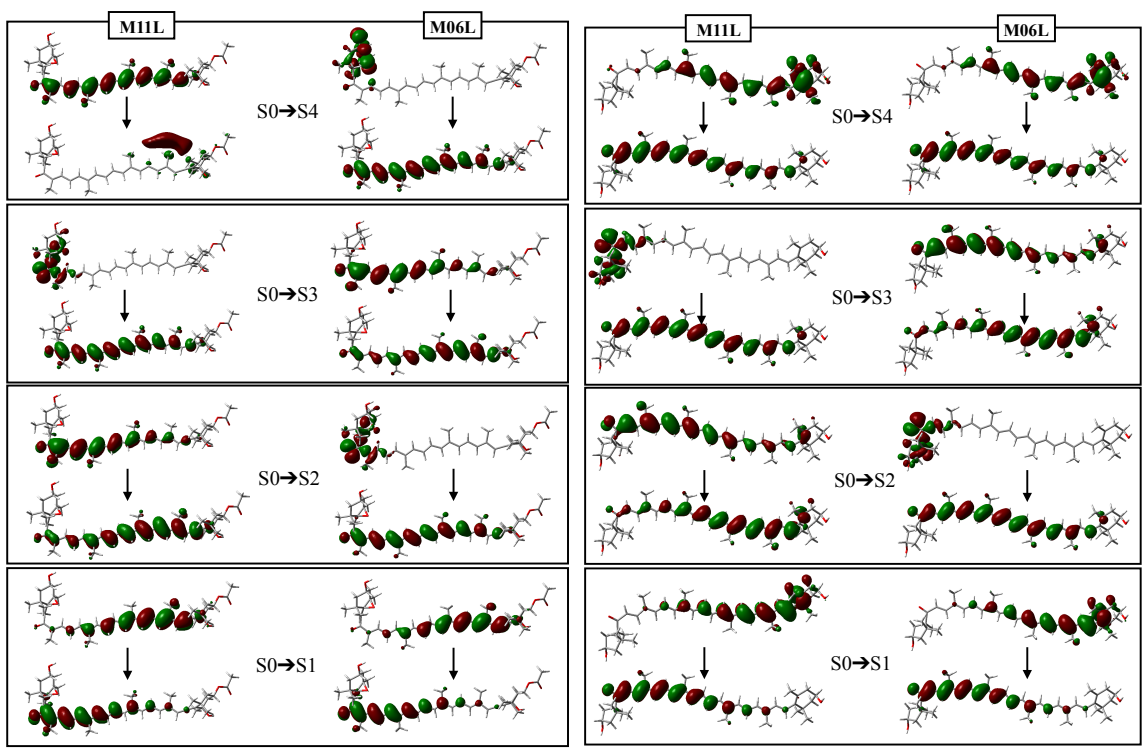

**Figure 10**. Comparison of dominant NTO pairs for the four lowest excitations of Fucoxanthin (left) and Capsanthin (right) obtained at M11L and M06L level.

Considering the excited-state description based on the M06L functional, we have to recall that an exchange between $S_2$ and $S_3$ was found for both Fucoxanthin and Capsanthin on the basis of the oscillator strengths reported in **Figure 5**. This becomes evident now in **Figure 10**: in both systems, the $S_2$ state assumes a clear n$\pi$* character while the bright $\pi\pi$* becomes $S_3$. By contrast, with M11L the description of the lowest transition of both systems and the fourth transition of Capsanthin are in good agreement with the DFT/MRCI reference data. Finally, the $S_0 \rightarrow S_4$, transition of Fucoxanthin is strangely described by M11L as being characterized by a Rydberg-like NTO.

Considering the role of the carbonyl group, the DFT/MRCI shows that it hardly influences neither the transition energy of the $\pi\pi$*

transitions nor their main excitation character (see **Figure 9**). The density difference located at the C=O moiety remains rather constant for all low-lying $\pi\pi^*$ transitions. In contrast, changing the conjugation pattern of the C=O moiety from cis to trans leads to an energetic lowering of the $n\pi^*$ transition bringing it closer to the lowest-lying $\pi\pi^*$ transitions which instead remain rather constant. This can be seen best from a comparison of the $\pi\pi^*$ transition energies of the Echinenone-TC and Echinenone-CT isomers which only differ in the connectivity of the C=O moiety (see **Figure 4**) and exhibit nearly identical transition energies to the $S_1$ and $S_2$ states.

## 4. Conclusions

We have studied the low-lying singlet excited states of a set of carbonyl containing carotenoids (Echinenone, Canthaxanthin, Astaxanthin, Fucoxanthin, Capsanthin and Capsorubin) by using DFT/MRCI. The analysis of the different transitions pointed out an important role of the C=O moiety: its presence leads to low-lying $n\pi^*$ states which may be close to the $\pi\pi^*$ states and thus influence the photophysics of this particular class of carotenoids. To which extent the $n\pi^*$ states become closer to the $\pi\pi^*$ ones depend on the length of the carotenoid. The smaller it is, the closer the $n\pi^*$ state(s) are to the (bright) $\pi\pi^*$ state. It is therefore tempting to suggest in this context a possible connection between these low-lying $n\pi^*$ states with the experimentally evidenced ICT state appearing in polar solvents for "short" carbonyl-containing carotenoids.[5,7,9] The attribution of such state is often subject to controversy: it can be a mixing of the lowest two excited singlet states,[67] a state that is not accessible by a vertical transition (or in any case a state which allows spontaneous florescence to the ground state only for geometries far from the Franck-Condon ones),[68] while more recently a direct connection between the ICT and the $n\pi^*$ state has been proposed.[12] Our multireference calculations show that indeed there are such states, corresponding to $n\pi^*$ states and that decreasing the polyene chain their energy decreases following the experimentally observed trend.

The DFT/MRCI results were also used to identify the "best" DFT functional (or class of functionals) to be used in combination with TDA-DFT calculations. Even if such an approach cannot account for the double excitation character of the different states -- in particular the $S_1$ state -- we found that the meta-GGA M11L and MN12L functionals provide correct transition energies and ordering for all the investigated carotenoids although the $S_1$ oscillator strengths are overestimated. These results suggest that TDA-DFT in combination with properly selected functionals can serve as an efficient computational tool to study the photophysics of carotenoids within their natural environment and beyond the limit of the Franck-Condon region.

## Acknowledgments

SK thanks C. M. Marian (U Düsseldorf) for providing him access to the reparametrized DFT/MRCI code and for fruitful discussions on the importance of $n\pi^*$ states in carotenoids.